\documentclass[journal]{IEEEtran} 
\usepackage{graphicx}
\graphicspath{ {./Figure/} }
\usepackage{array}
\usepackage{amssymb} 
\usepackage{amsmath} 
\usepackage[linesnumbered,ruled,vlined]{algorithm2e}
\usepackage{algpseudocode}
\newdimen\origiwspc%
\origiwspc=\fontdimen2\font

\usepackage{tikz,xcolor,hyperref}
\definecolor{lime}{HTML}{A6CE39}
\DeclareRobustCommand{\orcidicon}{%
	\begin{tikzpicture}
	\draw[lime, fill=lime] (0,0) 
	circle [radius=0.16] 
	node[white] {{\fontfamily{qag}\selectfont \tiny ID}};
	\draw[white, fill=white] (-0.0625,0.095) 
	circle [radius=0.007];
	\end{tikzpicture}
	\hspace{-2mm}
}

\foreach \x in {A, ..., Z}{%
	\expandafter\xdef\csname orcid\x\endcsname{\noexpand\href{https://orcid.org/\csname orcidauthor\x\endcsname}{\noexpand\orcidicon}}
}

\begin{document}
	\title{\huge Block Deep Neural Network-Based Signal Detector \\for Generalized Spatial Modulation}	
	\author{
		\IEEEauthorblockN{
			Hasan Albinsaid\orcidA{},~\IEEEmembership{Student Member,~IEEE},
			Keshav Singh\orcidB{},~\IEEEmembership{Member,~IEEE},
			Sudip Biswas\orcidC{},~\IEEEmembership{Member,~IEEE},
			Chih-Peng Li\orcidD{},~\IEEEmembership{Fellow,~IEEE},  and
			Mohamed-Slim Alouini\orcidE{},~\IEEEmembership{Fellow,~IEEE}
		}\vspace{-1.75em}
			\thanks{Manuscript received July 18, 2020; accepted August 6, 2020; updated May, 2021. This work was supported by the Ministry of Science and Technology of Taiwan under grants MOST 108-2218-E-110-014 and MOST 109-2218-E-110-006. The work of Dr. Sudip Biswas was supported by TEQUIP III of IIIT Guwahati. The associate editor coordinating the review of this letter and approving it for publication was G. Zheng. (Corresponding author: Chih-Peng Li.)}
			\thanks{Hasan Albinsaid, Keshav Singh, and Chih-Peng Li are with the Institute of Communications Engineering, National Sun Yat-sen University, Kaohsiung 80424, Taiwan (e-mail: hasan@g-mail.nsysu.edu.tw; keshav.singh@mail.nsysu.edu.tw; cpli@mail.nsysu.edu.tw).}
			\thanks{Sudip Biswas is with the Department of ECE, Indian Institute of Information Technology Guwahati, Guwahati 781015, India (e-mail: sudip.biswas@ieee.org).}
			\thanks{Mohamed-Slim Alouini is with the CEMSE Division, King Abdullah University of Science and Technology (KAUST), Thuwal 23955, Saudi Arabia (e-mail: slim.alouini@kaust.edu.sa).}		
			\thanks{Digital Object Identifier 10.1109/LCOMM.2020.3015810}
		}	
\markboth{IEEE COMMUNICATIONS LETTERS}%
{Shell \MakeLowercase{\textit{et al.}}: Bare Demo of IEEEtran.cls for IEEE Communications Society Journals}

	\IEEEtitleabstractindextext{%
		\begin{abstract}
			\fontdimen2\font=1ex
			Generalized Spatial Modulation (GSM) is being considered for high capacity and energy-efficient networks of the future. However, signal detection due to inter channel interference among the active antennas is a challenge in GSM systems and is the focus of this letter. Specifically, we explore the feasibility of using deep neural networks (DNN) for signal detection in GSM. In particular, we propose a block DNN (B-DNN) based architecture, where the active antennas and their transmitted constellation symbols are detected by smaller sub-DNNs. After $N$-ordinary DNN detection, the Euclidean distance-based soft constellation algorithm is implemented. The proposed B-DNN detector achieves a BER performance that is superior to traditional block zero-forcing (B-ZF) and block minimum mean-squared error (B-MMSE) detection schemes and similar to that of classical maximum likelihood (ML) detector. Further, the proposed method requires less computation time and is more accurate than alternative conventional numerical methods.
		\end{abstract}
	}
	
	\maketitle
	\IEEEoverridecommandlockouts
	\IEEEpubid{
		\begin{minipage}{\textwidth}\ \\ \\ \\ \\[12pt]
			\centering
		1558-2558~\copyright~2020 IEEE. Personal use is permitted, but republication redistribution requires IEEE permission. \\ See https://www.ieee.org/publications/rights/index.html for more information.
	\end{minipage}
	}
	\IEEEdisplaynontitleabstractindextext
	\IEEEpeerreviewmaketitle
	
	\vspace{-1.0em}
	\section{Introduction}
	\IEEEPARstart{I}{ncreasing} the number of antennas at the transmitter and receiver is a common trend in current wireless systems, whereby spatial multiplexing is utilized to achieve the demands of high transmission rates. However, such a technique requires plenty of radio frequency (RF) chains that not only add towards hardware cost and complexity~\cite{rf}, but also increase the power consumption of systems. Accordingly, to overcome these problems, Spatial Modulation (SM) was proposed, whereby at any instant of time only a single antenna is active and a block of any number of information bits is mapped into two constellation points, one each in the signal and spatial domain. 
	
	\fontdimen2\font=0.9ex
	However, the benefits of SM come at the cost of reduced data rates, when compared to current state-of-the-art multiple-input multiple-output (MIMO) systems. Hence, to overcome the limitations of SM, generalized SM (GSM) was proposed in \cite{gsm0}, whereby more than one antenna is active at any time slot. However, this introduces the issue of multi-antenna interference, which needs to be mitigated at the receiver through a detection algorithm. Nevertheless, with appropriate detection techniques it is possible for a GSM system to achieve higher spectral efficiency and lower bit error rate than spatial multiplexing systems by utilizing only a fraction of the available RF chains~\cite{Chen2018,Xiao2014}. Based on the above, in this work, we explore the possibility of using deep neural networks (DNN) for signal detection in GSM. Deep learning (DL) has reformed the way we formulate and optimize problems in several areas including
	image recognition, natural language processing, and speech recognition. Recently, DL has made significant inroads in various fields of wireless communications, like channel coding \cite{dnn0}, antenna selection, modulation classification \cite{dnn2}, etc. More importantly, the solutions achieved through DL have outperformed existing classical techniques. Accordingly, in this letter we design a DNN-based signal detector for a GSM MIMO system, with specific distinctions from existing works, which are noted below:
	
	\begin{itemize}
		\fontdimen2\font=0.75ex
		\item We adopt the concept of feature vector generator (FVG) in data pre-processing in order to convert the complex-valued IQ raw data into a clean data set. This process speeds up and enhances performance of the symbol classification process. Next, we propose a novel block DNN (B-DNN) architecture, wherein the active antennas and their transmitted constellation symbols are detected by smaller sub-DNNs. 
		\item Through  numerical results, we show that the BER performance of the proposed B-DNN based signal detector for GSM is much better than the traditional block zero-forcing (B-ZF) and block minimum mean-squared error (B-MMSE) detection schemes and it's performance is similar to classical maximum likelihood (ML) detector. 
	\end{itemize}
	\section{GSM-MIMO and Conventional Detection}
	\subsection{System Framework}
	We consider a GSM MIMO system with $N_t$ and $N_r$ transmitting and receiving antennas, respectively, with $N_r<N_t$ \footnote{For $N_r \geq N_t$, the simple MRC algorithm is able to provide optimal performance.}. In this system, only $N_p(2\leq N_p\ll N_t)$ transmit antennas are activated at any particular time slot. Therefore, the total combinations when choosing $N_p$ transmit antennas out of $N_t$ is  given by $N_t \choose N_p$ possible transmit antenna combinations (TACs), where $N_t \choose N_p$ represents the binomial coefficient. Among those TACs, only  $N=2^{\lfloor\log_2{N_t \choose N_p}\rfloor}$ TACs are permitted and the remaining combinations are considered illegitimate, where $\lfloor\cdot\rfloor$ denotes the floor operation.
	
	Next, the information bits are divided into two parts in each time slot, i.e., the TACs modulated bits and symbols modulated bits by quadrature amplitude modulation (QAM). \textit{N} combinations are chosen to convey $\log_2{N}$ TACs modulated bits and $N_p$ of \textit{M}-QAM symbol modulation convey $N_p\log_2{M}$ symbol modulated bits. Since we have $N_p$ active antennas and remaining $N_t-N_p$ are silent, the transmit vector $\mathbf{x}$ can be expressed as $\mathbf{x} = \lbrack \ldots,0,s_1,0,\dots,0,s_2,0,\dots,0,s_{N_p},0,\ldots \rbrack^{T}$, where the symbols $s_1,s_2,\dots,s_{N_p}\in\mathcal{S}$ and $\mathcal{S}$ is the constellation set of \textit{M}-QAM. As a result, $B=\log_2{N}+N_p\log_2{M}$ bits of information can be transmitted in each time slot.
	
	Let $\mathbf{H} \in \mathbb{C}^{N_r \times N_t}$ denote a quasi-static flat fading MIMO channel matrix, whose entries follow a complex Gaussian distribution $\mathcal{CN}(\mathbf{0,1})$. Then, the received signal $\mathbf{y}\in\mathbb{C}^{N_r\times1}$ can be formulated as
	\begin{equation} \label{eq_y}
	\mathbf{y}=\mathbf{Hx}+\mathbf{n}=\sum\nolimits_{k=1}^{N_p}\mathbf{h}_{i_k}s_{i_k}+\mathbf{n}=\mathbf{H}_I\mathbf{s}+\mathbf{n},
	\end{equation}
	where $\mathbf{n}\in\mathbb{C}^{N_r\times1}$ is the additive noise vector following complex Gaussian distribution $\mathcal{CN}(\mathbf{0},\sigma^2\mathbf{I})$, $\mathbf{h}_k$ is the \textit{k}-th column of $\mathbf{H}$, and $\mathbf{H}_I=(\mathbf{h}_{i_1},\dots,\mathbf{h}_{i_{N_p}})$ is the sub-matrix of $\mathbf{H}$ corresponding to the combination set $I$.
	\subsection{Conventional Detection}
	\subsubsection{ML Detector Schemes}
	The ML detector is an optimal detector commonly used in MIMO systems and it can be formulated as 
	\begin{equation} \label{eq_ML}
	(\hat{I},\hat{s})= \mbox{arg} \min_{I\in\mathbb{I},s\in\mathbb{S}} {\| \mathbf{y-\mathbf{H}_\mathit{I} \mathbf{s}}\|}_F^{2}
	\end{equation}
	where $\mathbb{I}=\{I_1,I_2,\dots,I_N\}$, $I_i$ with $i \in \{1,2,\dots,N\}$ is the set of illegitimate TACs, and $\mathbb{S}=\mathcal{S}^{N_p\times1}$ is the set of $N_p$-dimensional symbol vectors. Because the ML detection algorithm jointly detects the activated antennas and constellation points by exhaustive search from all possible transmitted signal vector, it causes high decoding complexity at the receiver.
	
	\subsubsection{Linear Detector Schemes}
	Since complexity of ML detection increases exponentially with the number of transmit antennas and modulation levels, some low-complexity linear detection schemes like zero-forcing (ZF) and minimum mean-squared error (MMSE) detection schemes have also been used in literature. The  ZF detection scheme is given by
	\begin{equation} \label{eq_ZF}
	\hat{\mathbf{x}}_{ZF}=(\mathbf{H}^{H}\mathbf{H})^{-1}\mathbf{H}^{H}\mathbf{y},
	\end{equation}
	where $\mathbf{H}^{H}$ is the conjugate transpose of $\mathbf{H}$. However, ZF detection has the issue of noise amplification with additive white Gaussian noise.
	In this regard, another low-complexity linear detection scheme, MMSE detection gives a better BER performance in comparison to ZF detection because it takes into consideration the noise level. The MMSE detection scheme is given by
	\begin{equation} \label{eq_MMSE}
	\hat{\mathbf{x}}_{MMSE}=(\mathbf{H}^{H}\mathbf{H}+\sigma^2\mathbf{I})^{-1}\mathbf{H}^{H}\mathbf{y}.
	\end{equation}
Note that the above linear detection schemes require more number of receive antennas than transmit antennas. 
	
	\subsubsection{Block Linear Detector Scheme}
	The linear detection schemes are supposed to solve the inverse operation for \eqref{eq_ZF} and \eqref{eq_MMSE} that have the channel matrix $\mathbf{H}$ with size $N_r \times N_t$. Therefore, we should solve an inverse of a $N_t \times N_t$ dimension matrix, and $N_r$ should at least be equal to $N_t$ to guarantee that $\mathbf{H}^{H}\mathbf{H}$ is of full rank. Since in GSM we have $N_r<N_t$  and $N_p$ antennas out of $N_t$ is activated, considering that $N_r \geq N_p$, we can simply apply \eqref{eq_ZF} and \eqref{eq_MMSE} by considering only the active antenna columns of channel matrix $\mathbf{H}_\mathit{I}$ instead of all the columns of the channel matrix $\mathbf{H}$. Therefore, we will have $N$ solutions of estimated transmit signal $\hat{\mathbf{s}}_{I}$ for ZF and MMSE detection. To get the final solution of estimated transmit signal we can look for the $\mathit{I}$-th euclidean distance between the received signal and multiply the channel matrix with \textit{I}-th estimated transmit signal $\hat{\mathbf{s}}_{I}$ as
\begin{equation} \label{eq_BZF}
	(\hat{I})= \mbox{arg} \min_{I\in\mathbb{I}} {\| \mathbf{y}-\mathbf{H}_{I}\hat{\mathbf{s}}_{I}\|}_F^{2}.
	\end{equation}
	The above block linear detection scheme performs better in terms of BER than the previously mentioned detection schemes, but its complexity increases with increasing number of transmit antennas.

	\section{Proposed GSM Block-DNN Detection}
	\begin{figure*}[ht]
		\centering
		\includegraphics[width=0.9\textwidth]{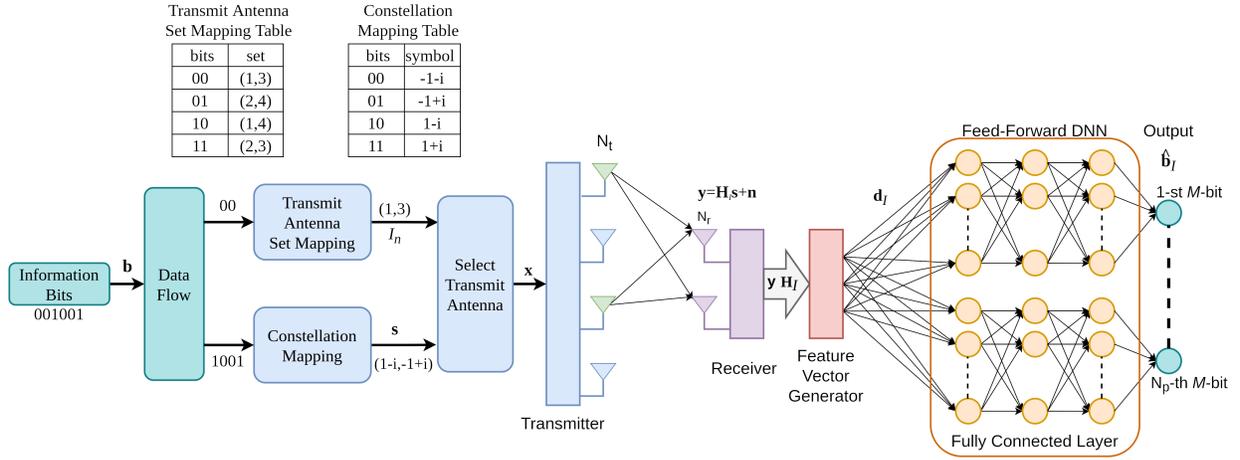}\vspace{-0.75em}
		\caption{Block diagram of Block-DNN signal detector for GSM transmitter and receiver.}
		\label{fig:gsm}\vspace{-0.75em}
	\end{figure*}
	Fig. \ref{fig:gsm} shows the block diagram of the proposed GSM transceiver. The Block-DNN detector mainly consists of two parts: FVG and feed-forward DNN.
	\vspace{-1.50em}
	\subsection{Data pre-processing}
	According to \cite{dnn1}, the key towards achieving better performance in a deep learning algorithm is the data pre-processing. Data pre-processing is the transformations applied to input data before feeding it into the DNN model. In particular, the aim is to convert the raw data into a clean data set. 
	\subsubsection{Raw data}
	The received signal vector \eqref{eq_y} is received by $N_r$ antennas. Assuming perfect channel state information at receiver (CSIR), and the fact that combination of TACs is known at the receiver, we have a vector $\mathbf{y}$ and $N$ combination of active antennas column of channel matrix $\mathbf{H}_\mathit{I}$.
	\subsubsection{Feature vector generator}
	Since our raw data is multiple vector/matrix with complex-valued IQ elements, it should be converted into vector data. Here we introduce a separate FVG (SFVG). The idea of SFVG is to separate the raw complex-valued IQ data into real vector. For example, if there is a channel matrix $\mathbf{H}$ which will be extracted by using SFVG, it can be expressed as\vspace{-0.5em}
	\begin{equation}
	\begin{split}
	\label{eq_SFVG}
	\mathbf{f}_{SFVG}(\mathbf{H})=&[|\Re(h_{1,1})|,|\Im(h_{1,1})|,\\
	&\dots,|\Re(h_{a,b})|,|\Im(h_{a,b})|,\\
	&\dots,|\Re(h_{N_r,N_t})|,|\Im(h_{N_r,N_t})|]^{T}\,.
	\end{split}
	\end{equation}
	\subsubsection{Final vector input}
	Let $\mathbf{D}^{(j)}$ be the $\mathit{j}$-th entry of $N$ final vector data set input that can be expressed as $\mathbf{D}^{(j)}=[\mathbf{d}_1,\dots,\mathbf{d}_i,\dots,\mathbf{d}_N]$, where $j \in\{1,2,\dots,N_s\}$ is the time slot index of block information stream. Now, as the provided raw data is a vector $\mathbf{y}$ and $N$ combination of active antennas column of channel matrix $\mathbf{H}_\mathit{I}$, so each $\mathbf{d}_i$ is given by\vspace{-0.5em}
	\begin{equation}
	\label{eq_di}
	\mathbf{d}_i=[\mathbf{f}_{SFVG}(\mathbf{y}^{(j)})^{T}, \mathbf{f}_{SFVG}(\mathbf{H}^{(j)}_{\mathbf{\mathit{I}}_{i}})^{T}]^{T},
	\end{equation}
	where $\mathbf{d}_i \in \mathbb{R}^{(2N_r+2N_rN_p)\times1}$.
	
	\subsection{Feed-forward DNN parameters and training}
	$L$ fully connected layers with $L-1$ hidden layers is considered for decoding each of the active transmit antenna of transmitter. Table. \ref{tab:table1} shows the number of layers (we have $\delta_l$ nodes in the $l$-th layer) and parameters and their corresponding values for the proposed DNN.
	
\begin{figure}[t!]
		\centering
		\includegraphics[width=0.35\textwidth]{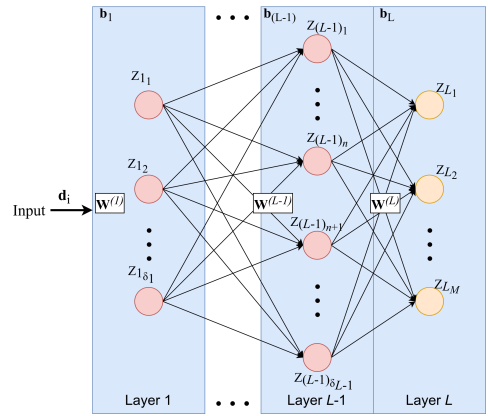}
		\caption{Layer parameter of fully  connected layer.}\vspace{-1.0em}
		\label{fig:layer}
	\end{figure}

	\begin{table}[t!]
		\renewcommand{\arraystretch}{1.0}
		\begin{center}
			\vspace{-0.5em}
			\caption{Network and training parameters}
			\label{tab:table1}
			\vspace{-0.5em}
			\begin{tabular}{|m{1.6cm}|>{\centering\arraybackslash}m{1.8cm}||m{1.9cm}|>{\centering\arraybackslash}m{1.55cm}|}
				\hline
				\textbf{Parameters} & \textbf{Value} & \textbf{Parameters} & \textbf{Value}\\ 
				\hline
				\hline
				Input nodes & $2(N_r+N_rN_p)$ &  Learning rate & 0.005 \\ 
				\hline
				Hidden layer & 3 & Number of training set & 15.000.000 \\ 
				\hline
				Output nodes & $M$ & Number of validation set & 5.000.000 \\ 
				\hline
				Hidden layer activation & ReLu & Epoch & 50 \\ 
				\hline
				Output layer activation & Softmax & BPSK hidden nodes & 128- 64-32\\ 
				\hline
				Loss function & Cross-entropy & QPSK hidden nodes & 256-128-64 \\ 
				\hline
				Optimizer & SGD & 16-QAM hidden nodes & 512-256-128\\ %
				\hline
			\end{tabular}
		\end{center} \vspace{-2.5em}
	\end{table}
	\fontdimen2\font=0.8ex
	We use $\lambda$ to denote the set of all the parameters of DNN, $\lambda=\{\lambda_1,\lambda_2,\dots,\lambda_L\}$. The set of $l$-th layer parameter is denoted by $\lambda_l=\{\textbf{W}^{(l)},\textbf{b}_l\}$. Accordingly, the $l$-th layer is given by
	\begin{equation}
	\label{eq_Z}
	\mathbf{Z}_l=\sigma({\mathbf{W}^{(l)}}^{T}\mathbf{Z}_{l-1}+\mathbf{b}_l),
	\end{equation}
	\fontdimen2\font=0.76ex
	where $\sigma(\cdot)$ is an activation function, $\textbf{W}^{(l)} \in \mathbb{R}^{\delta_{l-1}\times \delta_{l}}$ is the weight matrix and $\textbf{b}_{l} \in \mathbb{R}^{\delta_{l}\times 1}$ is the bias vector. At each layer except the last, rectified linear unit (ReLU) function is used, with $\sigma(x)=max(0,x)$ as the activation function. The gradient of this function is always a single value, either 0 or 1, which ensures that the size of the gradients is not exponentially reduced as we back-propagate through many layers. ReLU learns quickly in DNN, allowing training of a deep supervised network without unsupervised pre-training \cite{glorot2011deep}. In the last layer, softmax function \cite{dnn2} is used to map the output in the range [0,1]. The input and output mapping of $L$-layer of DNN series functions depicted in Fig.~\ref{fig:layer} are expressed by\vspace{-0.5em}
	\begin{multline}
	\label{eq_ZL}
	\mathbf{Z}_L=\sigma\big(\mathbf{W}^{(L)}(\sigma\big(\mathbf{W}^{(L-1)}(\dots \\
	\sigma\big(\mathbf{W}^{(1)}\mathbf{Z}_0+\mathbf{b}_1\big)\dots)+\mathbf{b}_{L-1}\big))+\mathbf{b}_{L}\big)\,,
	\end{multline}
	where $\mathbf{Z}_0$ is equal to the final vector input $\mathbf{d}_i$.
	
	\vspace{-0.25em}
	Categorical cross entropy is applied to look for the cost function between the true data and prediction data, through which we can get the parameter to optimize our network. 
	Let $\textbf{Z}_{T}$ be one-hot vector for labeling the supervised training which will be compared to the results of the prediction $\mathbf{Z}_L$. Then, the cross-entropy cost function is expressed \vspace{-0.50em}	
	\begin{equation}
	\label{eq_Lost}
	\hat{\mathit{L}}(\mathbf{Z}_{T},\mathbf{Z}_L)=-\sum\nolimits_{n=1}^{M}Z_{T_n}\log(Z_{L_n}).
	\end{equation}
	
	\begin{figure*}[t!]
		\centering
		\includegraphics[trim=70 20 15 15,width=0.95\textwidth]{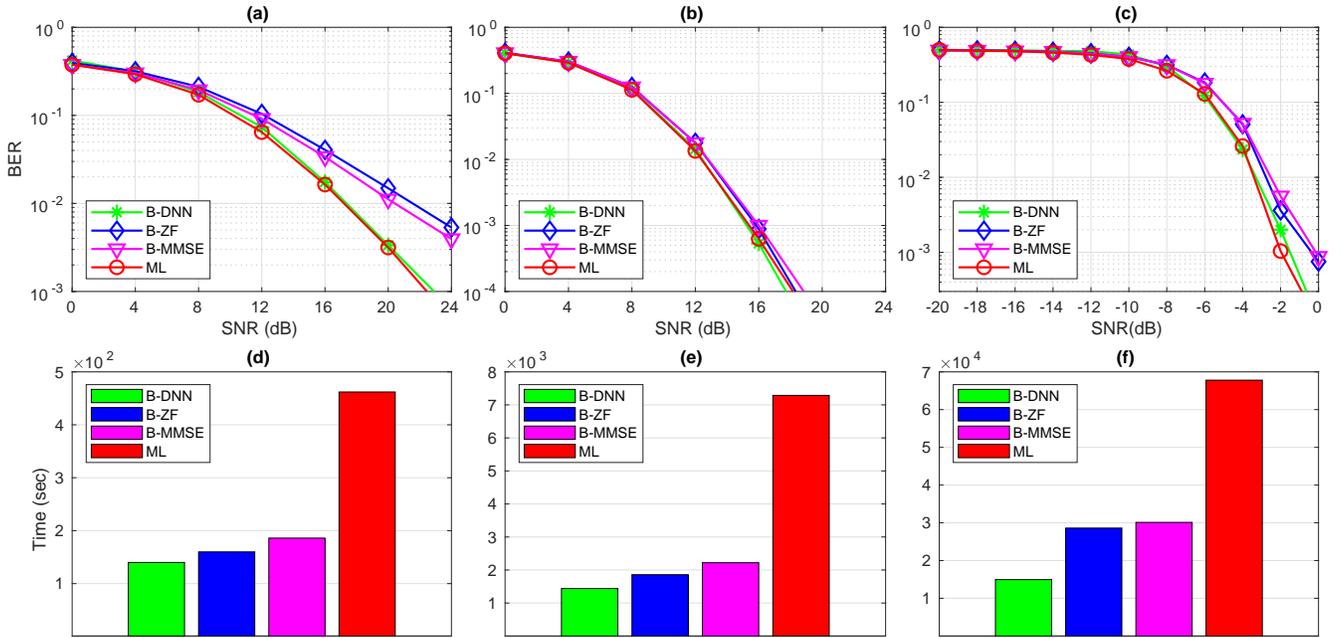}\vspace{-0.75em}
		\caption{BER and time complexity comparison for various detectors under QPSK modulation with (a,d) $N_s=100000$, $ N_t=4 $, $N_p=2$, $N_r=2$, (b,e) $N_s=100000$, $ N_t=16 $, $N_p=2$, $N_r=4$, (c,f) $N_s=5000$, $ N_t=128 $, $N_p=2$, $N_r=64$.}\vspace{-1.0em}
		\label{fig:result}
	\end{figure*} 
\noindent Because the difference between $Z_{T_n}$ and $Z_{L_n}$ should be as small as possible, we want to make an effort to minimize it. This is done through stochastic gradient descent (SGD) optimization by subtracting or adding the old weight value to the learning rate of the gradient value that we have, where the learning rate $\eta$ is a hyperparameter in the range between $0$ and $1$. SGD  iteratively updates the values over time by using the value of the gradient, as shown in the equation below.\vspace{-0.5em}
	\begin{equation}
	\label{eq_SGD}
	\mathbf{W}_{t+1}^{(l)}=\mathbf{W}_{t}^{(l)}-\eta\frac{\partial\hat{\mathit{L}}(\mathbf{Z}_{T},\mathbf{Z}_L)}{\partial\mathbf{W}_{t}^{(l)}}.
	\end{equation}
	The weight update will be repeated continuously until the value of the cost-function over time has saturated. When the optimal weight is achieved, predictions can be done by using the trained feed-forward DNN.
	
	During the training process, the training inputs and labels are required. The training inputs are generated from the received signal $\mathbf{y}$ (without noise) and the channel matrix $\mathbf{H}$, which can be expressed as: $\mathbf{d}_{i}=[$ $\mathbf{f}_{SFVG}(\mathbf{y}^{(j)})^{T}$, $ \mathbf{f}_{SFVG}(\mathbf{H}^{(j)}_{\mathbf{\mathit{I}}_{i}})^{T}$ $]^{T}$ $\!=\![$ $\mathbf{f}_{SFVG}(\mathbf{H}^{(j)}\mathbf{x}^{(j)})^{T}$, $ \mathbf{f}_{SFVG}(\mathbf{H}^{(j)}_{\mathbf{\mathit{I}}_{i}})^{T}$ $]^{T}$. The labels are generated by utilizing one-hot vector of size $M$ from transmitted symbol to represent \textit{M}-QAM symbol constellation.
	\vspace{-1.0em}
	\subsection{Prediction}
	Since in the $j$-th time slot of prediction we have $N$ input vectors $\mathbf{d}_i$ and each of it produces $N_p$ output vectors $\mathbf{Z}_L \in \mathbb{R}_{+}^{M\times1}$ for the $k$-th active transmit antenna, we can get the prediction of transmitted symbol $\hat{s}_{i_k}=\mathcal{S}_{\hat{n}}$, and the index $\hat{n}$ is given by\vspace{-0.5em}
	\begin{equation}
	\label{eq_indexout}
	(\hat{n})=\operatorname*{arg\,max}_{n\in \{1,\dots,M\}}(\mathbf{Z}_{L_n}).
	\end{equation}
	Therefore, we have $N$ output predicted symbol vector $\hat{\mathbf{s}}_i$, where $\hat{\mathbf{s}}_i=[\hat{s}_{i_1},\dots,\hat{s}_{i_k},\dots,\hat{s}_{i_{N_p}}]$. We can now look for the solution by calculating the euclidean distance between the received signal and the predicted symbol, which can be expressed as\vspace{-0.5em}
	\begin{equation} \label{eq_euclideanY}
	(\hat{i})= \mbox{arg} \min_{i\in\{1,\dots,N\}} {\| \mathbf{y}^{(j)}-\mathbf{H}^{(j)}_{\mathbf{\mathit{I}}_{i}}\hat{\mathbf{s}}^{(j)}_{i}\|}_F^{2}.
	\end{equation}
	By using the minimum distance index $\hat{i}$, we take the bit mapping of symbol $\hat{\mathbf{s}}_{\hat{i}}$. Finally, we obtain the demodulation of information vector $\hat{\mathbf{b}}$ as the output of the block-DNN GSM detector.	
	\SetKwInput{KwInput}{Input}                
	\SetKwInput{KwOutput}{Output}              
	\SetKwInput{KwInpDNN}{Input size DNN}
	\SetKwInput{KwHdAct}{Hiden layer activation}
	\SetKwInput{KwOutAct}{Output layer activation}
	\SetKwInput{KwOutDNN}{Output size DNN}
	\begin{algorithm}[t!]
		\KwInput{$\mathbf{y}^{(j)}$, $\mathbf{H}^{(j)}$,  $\mathbf{I}$, $N$, $N_s$ }
		
		\For{$j\gets1$ to $N_s$ by $1$}{
			\For{$i\gets1$ to $N$ by $1$}{
				$\mathbf{d}_{i}=[$ $\mathbf{f}_{SFVG}(\mathbf{y}^{(j)})^{T}$, $ \mathbf{f}_{SFVG}(\mathbf{H}^{(j)}_{\mathbf{\mathit{I}}_{i}})^{T}$ $]^{T}$
			}
			$\mathbf{D}^{(j)}=[\mathbf{d}_1,\mathbf{d}_2,\dots,\mathbf{d}_N]$
		}
		
		$\mathbf{D}=[\mathbf{D}^{(1)},\mathbf{D}^{(2)},\dots,\mathbf{D}^{(N_s)}]$
		
		$\hat{\mathbf{S}}=\mathit{DNN}(\mathbf{D})$; where $\hat{\mathbf{S}}=[\hat{\mathbf{S}}^{(1)},\hat{\mathbf{S}}^{(2)},\dots,\hat{\mathbf{S}}^{(N_s)}]$, $\hat{\mathbf{S}}^{(j)}=[\hat{\mathbf{s}}_{1},\hat{\mathbf{s}}_{2},,\dots,\hat{\mathbf{s}}_{N}]$
		
		\For{$j\gets1$ to $N_s$ by $1$}{
			$(\hat{i})= \mbox{arg} \min_{i\in\{1,\dots,N\}} {\| \mathbf{y}^{(j)}-\mathbf{H}^{(j)}_{\mathbf{\mathit{I}}_{i}}\hat{\mathbf{s}}^{(j)}_{i}\|}_F^{2}$
			
			$\hat{\mathbf{b}}^{(j)}= $ bit mapping of $\hat{\mathbf{s}}_{\hat{i}}^{(j)}$
		}
		
		\KwOutput{Decoded bit $\hat{\mathbf{B}}=[\hat{\mathbf{b}}^{(1)},\hat{\mathbf{b}}^{(2)},\dots,\hat{\mathbf{b}}^{(N_s)}]$ }
		
		\caption{Proposed Block-DNN GSM Detector}
	\end{algorithm}
	\setlength{\textfloatsep}{0.1cm}
	\setlength{\floatsep}{0.2cm}
	
	\section{Simulation Results}
    In this section we analyse the performance and time complexity of the proposed B-DNN detectors for GSM systems through numerical simulations\footnote{The network was implemented in Tensorflow \cite{abadi2016tensorflow} and it was simulated on a standalone Ubuntu 20.04 PC with an AMD Ryzen 9 3950x CPU, NVIDIA GeForce RTX 2070 GPU, and 64 GB RAM.}.
    
    \begin{figure*}[t!]
    	\centering
    	\includegraphics[trim=2 0 0 2,clip,width=0.9\textwidth]{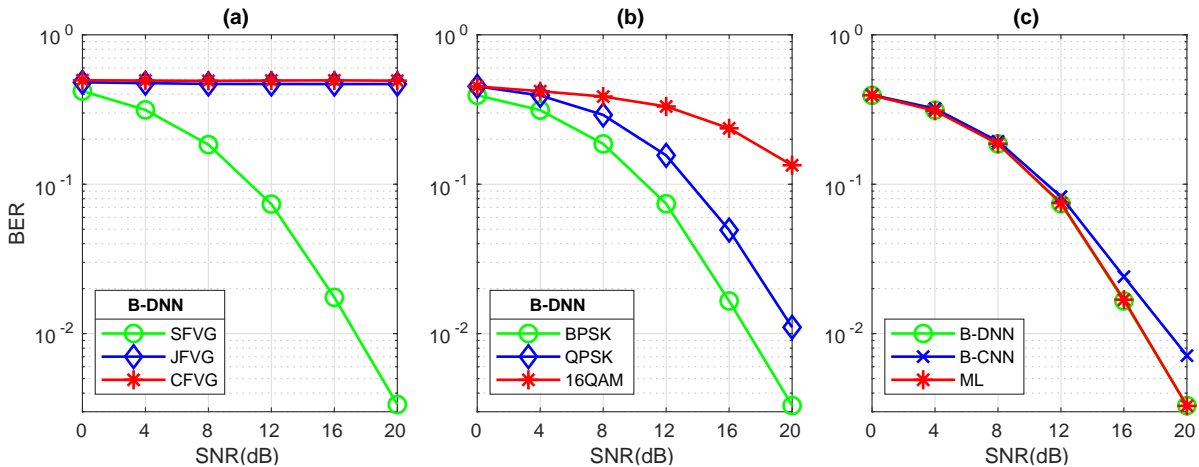}
    	\vspace{-0.5em}
    	\caption{{(a) BER comparison of proposed B-DNN with various FVG method in BPSK modulation}, (b) BER performance of proposed B-DNN for various modulation techniques, (c) BER comparison of B-DNN, B-CNN and ML in BPSK modulation. Here, we set $N_t=4$, $N_p=2$, $N_r=2$.}
    	\vspace{-0.75em}
    	\label{fig:berdnn}
    \end{figure*}
    \fontdimen2\font=0.5ex
    We begin by considering a GSM MIMO system with three different parameters that employ QPSK modulation. Fig. \ref{fig:result} shows the comparison between BER performance and computation time required for the proposed B-DNN detectors with respect to various conventional detectors. It can be seen that the BER performance for the scheme using the proposed B-DNN is quite close to that of the ML detector's performance. However, in Fig. \ref{fig:result}(a,d) it can be seen that the proposed B-DNN reduces the computation time by 69\%, 25\% and 13\% with respect to ML, B-MMSE, and B-ZF detectors, respectively. Further, in Fig. \ref{fig:result}(b,e) it can be seen that the proposed B-DNN reduces the computation time by 80\%, 35\% and 23\% when compared to ML, B-MMSE, and B-ZF detectors, respectively. Similarly, in Fig. \ref{fig:result}(c,f) the proposed B-DNN reduces the computation time by 78\%, 50\% and 48\% for ML, B-MMSE, and B-ZF detectors, respectively. 
    
	\fontdimen2\font=0.5ex
	Next, in Fig. \ref{fig:berdnn}(a) we show the BER performance of the proposed B-DNN detectors with various feature extraction methods, i.e., joint FVG (JFVG), conventional FVG (CFVG) and SFVG~\cite{fvg}. It can be seen that SFVG gives the best performance because it provides individual real and imaginary scalar values of the received signal and channel matrices. Fig. \ref{fig:berdnn}(b) on the other hand shows the BER performance of the proposed B-DNN detectors with respect to different modulation schemes, namely BPSK, QPSK and 16QAM modulation. As expected, the BER performance degrades with increasing order of modulation. 
    In Fig. \ref{fig:berdnn}(c) we compare the BER performance of the proposed B-DNN detector with respect to ML and a modified B-DNN. In particular, the modified B-DNN is designed by adding the following configuration (on top of the convolutional layer): [64 filter - max pooling - 128 filter - max pooling - 256 filter - max pooling] between FVG and feed-forward DNN (we call it block convolutional neural network (B-CNN)~\cite{o2016convolutional}). It can be seen that the performance of B-CNN is worse than that of B-DNN because B-CNN's learning of the feature signal is not in a serial form. As stated in \cite{o2016convolutional}, the feature learning methods of CNN has optimal performance only when the input is provided in a serial form.
    

Finally, we present the computational complexity of the proposed B-DNN detectors with respect to the various conventional detectors in terms of multiply-and-accumulate (MAC) operations in Table \ref{tab:table2}

	\begin{table}[t!]
		\renewcommand{\arraystretch}{1.3}
		\caption{Theoretical analysis of computational complexity\vspace{-1.50em}}
		\label{tab:table2}
		\begin{center}
			\begin{tabular}{|l|>{\centering\arraybackslash}m{6.3cm}|}
				\hline
				\textbf{Detector} & \textbf{Real-valued MAC}\\ 
				\hline
				\hline
				ML & $2^B(8N_rN_p+4N_r-1)$ \\ 
				\hline
				B-ZF & $N(4N_p^3+12N_p^2N_r+7N_p^2+6N_rN_p+6N_r-2N_p-1)$ \\ 
				\hline
				B-MMSE & $N(4N_p^3+12N_p^2N_r+7N_p^2+6N_rN_p+6N_r-1)$ \\ 
				\hline
				B-DNN & $NN_p((4N_rN_p+4N_r-1)\delta_1+\sum\limits_{k=1}^{k=L-1}\delta_{k+1}(2\delta_{k}-1))$ \\ 
				\hline
			\end{tabular}
		\end{center}\vspace{-0.50em}
	\end{table}
Since the computational complexity of B-DNN is primarily dependent on $\delta_l$, any change in the number of antennas in the transmitter or receiver doesn't have a greater impact on its complexity unlike conventional detectors.

	\section{Conclusion}
	\fontdimen2\font=\origiwspc
	In this letter, we proposed a B-DNN based detection scheme for a GSM system. Because of its fundamental ability to adequately learn the hidden interference plus noise models in practical receivers, the proposed B-DNN based detector achieves considerably better performance in terms of either BER or computation time when compared to standard detection techniques. In particular, through numerical results we verified that the BER performance of the proposed B-DNN scheme is better than B-ZF and B-MMSE detection schemes. Further, although the BER performance of the B-DNN does not right away outperform the classical ML detection technique, whereby the BER of both schemes almost overlap each other, the proposed technique comprehensively outperforms the ML scheme in terms of the required computation time. 
	
	\bibliographystyle{IEEEtran}
	
	\vspace{-0.75em}
	\bibliography{IEEEabrv,bibliography_ex}
	
\end{document}